\documentclass[twocolumn,nofootinbib,notitlepage]{revtex4-1}
\usepackage{epsfig,amsmath,slashed}

\newcommand{\di}{{\rm d}}

\def\wrho{{\widehat{\rho}}}

\newcommand{\tr}{{\rm tr}}

\newcommand{\omegav}{\boldsymbol{\omega}}

\newcommand{\be}{\begin{equation}}
\newcommand{\ee}{\end{equation}}                                                                               
\newcommand{\bea}{\begin{eqnarray}}
\newcommand{\eea}{\end{eqnarray}}                                                                               

\begin{document}

\title{Covariant statistical mechanics and the stress-energy tensor} 

\author{F. Becattini}\affiliation{Universit\`a di 
 Firenze and INFN Sezione di Firenze, Florence, Italy} 

\begin{abstract}
After recapitulating the covariant formalism of equilibrium statistical mechanics in
special relativity and extending it to the case of a non-vanishing spin tensor, we show 
that the relativistic stress-energy tensor at thermodynamical equilibrium 
can be obtained from a functional derivative of the partition function with respect to
the inverse temperature four-vector $\beta$. For usual thermodynamical equilibrium, the 
stress-energy tensor turns out to be the derivative of the relativistic 
thermodynamic potential current with respect to the four-vector $\beta$, i.e. 
$T^{\mu \nu} = - \partial \Phi^\mu/\partial \beta_\nu$. This formula establishes 
a relation between stress-energy tensor and entropy current at equilibrium 
possibly extendable to non-equilibrium hydrodynamics.
\end{abstract}

\maketitle


One of the best known formulae of statistical mechanics states that, in the
grand-canonical ensemble, the mean value of energy can be obtained as a 
derivative of the logarithm of the partition function with respect to the
inverse temperature. In terms of densities, the energy density is the derivative 
of the pressure with respect to the inverse temperature, i.e.:
$$
  \epsilon = -\frac{\partial (p/T)}{\partial (1/T)}\Big|_{\mu/T} 
$$
In relativity, the mean value of energy density is the time-time component of the
stress-energy tensor, so the question arises whether it is possible to obtain the 
full stress-energy tensor, and not just its time-time component, as a sort of derivative 
of the ``density" of partition function. In this work we provide an answer to this 
question and we will show that:
\be\label{main}
  T^{\mu \nu} = - \frac{\slashed \partial \Phi^\mu}{\slashed \partial \beta_\nu} 
  \Bigg|_{\rm eq}
\ee
that is the stress-energy tensor can be obtained as a variational Euler-Lagrange 
derivative (in a sense which will become clear later) at thermodynamical equilibrium of 
the relativistic generalization of the density of the thermodynamic potential $\log Z$, 
a vector current $\Phi^\mu$ which is linearly related to the entropy current.  

In other words, equation (\ref{main}) establishes a relation between the mean 
value of the stress-energy tensor and the relativistic entropy current at equilibrium.
If this relation could be extended to non-equilibrium situations, one would have 
a tool to determine entropy current from the expression of the stress-energy tensor,
which is one of the main problems of dissipative relativistic hydrodynamics \cite{israel1,
israel2} (for a recent discussion see also ref.~\cite{jensen}). The expression of 
entropy current in non-equilibrium is actually used as means to determine the 
structure of the stress-energy tensor itself and the generalization of the relation
(\ref{main}) would therefore be very important. 

Before proving eq.~(\ref{main}), we will recapitulate the fully covariant formulation 
of equilibrium in relativistic statistical mechanics, including the possibility of a
non-vanishing spin tensor, hence of a non-symmetric stress-energy tensor. We start 
with a brief summary of equilibrium thermodynamics of a systems with angular momentum, 
which macroscopically corresponds to a rigidly rotating fluid \cite{landau}.
A detailed discussion of this system can be found in refs.~\cite{becarot}. 


Thermodynamical equilibrium occurs when entropy is maximal. The maximization of
$S= -\tr \left( \wrho \log \wrho  \right)$ with the constraint of fixed, constant 
mean energy, charge and mean total angular momentum leads, as it is well known, to 
the density operator \cite{landau,balian}:
\be\label{densop}
  \wrho = \frac{1}{Z} \exp [ -\widehat H/T + \omega \widehat J_z/T  + 
   \mu \widehat Q/T]
\ee
where $Z = \tr(\exp [ -\widehat H/T + \omega \widehat J_z/T  + \mu \widehat Q/T])$.
The physical meaning of $T$ is that of temperature of an energy reservoir and $\omega$ 
that of an angular velocity of an angular momentum reservoir in contact with the system.
In a quantum relativistic field theory, the operators in eq.~(\ref{densop}) can be 
written as integrals over some space region $V$:
\bea\label{operat}
\!\!\!\!\!\!&& \widehat H = \int_V \di^3 x \; \widehat T^{00}(x) \qquad \qquad
 \widehat Q = \int_V \di^3 x \; \widehat j^{0}(x) \nonumber \\
\!\!\!\!\!\!&& \widehat J_z = \widehat J^{12} = \!\!\! \int_V \di^3 x 
 \left( x^1 \widehat T^{02}(x) - x^2 \widehat T^{01}(x) + \widehat{\cal S}^{0,12} (x) \right) 
\eea
$\widehat {\cal S}$ being the spin tensor. The operator (\ref{densop}) can be written 
in a fully covariant form. First define:
\be\label{betarot}
 \beta = (1/T) (1,\omegav \times {\bf x})  \qquad \qquad \xi = \mu/T
\ee 
with $\omegav = \omega {\bf \hat k}$ constant angular velocity vector directed along 
the $z$ axis. Then:
\be\label{omegat}
  \omega_{\lambda \nu} = \omega/T \left( \delta^1_\lambda \delta^2_\nu
  -\delta^2_\lambda \delta^1_\nu \right)
\ee
which is the acceleration tensor for a rigid rotation \cite{becarot}.
Finally, define the normal versor of the three-dimensional space-like hypersurface 
$V$ (embedded in Minkowski spacetime) appearing in the eq.~(\ref{operat}) as $\hat t$ 
and its measure $\di \Sigma_\mu \equiv \di^3 x \hat t_\mu$. Hence, we can rewrite 
(\ref{densop}) as:
$$
 \wrho = \frac{1}{Z} \exp \left[ \int_{V} \di\Sigma_\mu \; \left( 
  - \widehat T^{\mu \nu} \beta_\nu + \frac{1}{2} \widehat{\cal S}^{\mu,\lambda \nu} 
   \omega_{\lambda \nu} + \xi \widehat j^\mu \right) \right]
$$
The four-vector $\beta$ in eqs. (\ref{betarot}) is then, by construction, the inverse 
temperature four-vector and $1/\sqrt{\beta^2} = T/\sqrt{1-\|\omegav \times {\bf x}\|^2} 
\equiv T_0$ is the invariant temperature, i.e. the temperature measured by a thermometer 
moving with the rigid velocity field $\omegav \times {\bf x}$ with respect to the 
thermostat at temperature $T$ \cite{israel1,becarot}. The latter expression
does not fulfill yet the request of full covariance as it apparently depends on a
particular hypersurface $V$. In fact, if the divergence of the integrand vanishes 
and if its flux at the boundary of $V$, i.e. $\partial V$, also vanishes, the space-like 
hypersurface is arbitrary and the density operator can be finally written as:
\be\label{covdens2}
 \wrho = \frac{1}{Z} \exp \left[ \int_{\Sigma} \di\Sigma_\mu \; \left( 
  - \widehat T^{\mu \nu} \beta_\nu + \frac{1}{2} \widehat{\cal S}^{\mu,\lambda \nu} 
   \omega_{\lambda \nu} + \xi \widehat j^\mu \right) \right]
\ee
where $\Sigma$ is a general, arbitrary, space-like hypersurface bounded by the same 
$\partial V$. The covariant form (\ref{covdens2}) of the equilibrium statistical 
operator, to our knowledge, was first written down by Weldon \cite{weldon}; the above 
form generalizes his formula in that it includes a non-vanishing spin tensor, which 
is generally needed if the stress-energy tensor is not the Belinfante symmetrized 
stress-energy tensor \cite{hehl}. 

The two aforementioned conditions on the integrand also ensure the stationarity of 
the density operator with respect to any inertial frame, because $\wrho$ is unchanged 
in (\ref{covdens2}) going from the hypersurface $V(t)$ to $V(t+\Delta t)$. 
The divergence of the integrand reads:
\bea\label{div}
&& \partial_\mu \left( - \widehat T^{\mu \nu} 
  \beta_\nu + \frac{1}{2} \widehat{\cal S}^{\mu,\lambda \nu} \omega_{\lambda \nu} + 
  \xi \widehat j^\mu \right) = - \widehat T^{\mu \nu} \partial_\mu \beta_\nu \nonumber \\
&&  + \frac{1}{2} \omega_{\lambda \nu} 
  \partial_\mu \widehat{\cal S}^{\mu,\lambda \nu} + 
  \frac{1}{2} \widehat{\cal S}^{\mu,\lambda \nu} \partial_\mu \omega_{\lambda \nu} 
  + \widehat j^\mu \partial_\mu \xi
\eea
where we have taken into account the continuity equations of the stress-energy tensor 
and the current. As $\xi = \mu/T$ and $\omega_{\lambda\nu}$ (see eq.~\ref{omegat})
are constant, the above expression reduces to, after separating the symmetric and antisymmetric
part of the stress-energy tensor:
\bea\label{div2}
 &&  - \frac{1}{2}\widehat T^{\mu \nu}_S \left( \partial_\mu \beta_\nu + 
   \partial_\nu \beta_\mu \right)
  - \frac{1}{2}\widehat T^{\mu \nu}_A \left( \partial_\mu \beta_\nu - \partial_\nu \beta_\mu 
  \right) \nonumber \\
 &&  + \frac{1}{2} \omega_{\lambda \nu} \partial_\mu \widehat{\cal S}^{\mu,\lambda \nu} 
\eea
Because of the continuity equation of the angular momentum tensor:
\be
 \partial_\mu \widehat{\cal J}^{\mu,\lambda \nu} = 
 \partial_\mu \widehat{\cal S}^{\mu,\lambda \nu} + 2 {\widehat T}^{\lambda \nu}_A = 0
\ee  
the (\ref{div2}) vanishes if: 
\bea\label{equilib}
 &&  \partial_\mu \beta_\nu + \partial_\nu \beta_\mu = 0 \\ \nonumber
 &&  \omega_{\mu \nu} = - \frac{1}{2} \left(\partial_\mu \beta_\nu -
      \partial_\nu \beta_\mu \right)  
\eea
It can be readily checked that the four-vector $\beta$ and the tensor $\omega$ in 
eqs.~(\ref{betarot}) and (\ref{omegat}) fulfill both conditions above.
Hence, the divergence of the integrand does vanish:
\be\label{divvan}
  \partial_\mu \left( - \widehat T^{\mu \nu} 
  \beta_\nu + \frac{1}{2} \widehat{\cal S}^{\mu,\lambda \nu} \omega_{\lambda \nu} + 
  \xi \widehat j^\mu \right) = 0
\ee
The second condition, i.e the vanishing at the boundary:
\bea\label{boundary}
 && 0 = \int_{\rm boundary} \di \Sigma_\mu \left( -\widehat T^{\mu \nu} 
  \beta_\nu + \frac{1}{2}\widehat{\cal S}^{\mu,\lambda \nu} \omega_{\lambda \nu} + 
  \xi \widehat j^\mu \right) \nonumber \\
 &=& 
  \int_{t_0}^{t_1} \int_{\partial V} \di S \; n_\mu \left( - \widehat T^{\mu \nu} 
  \beta_\nu + \frac{1}{2}\widehat{\cal S}^{\mu,\lambda \nu} \omega_{\lambda \nu} + 
  \xi \widehat j^\mu \right)
\eea
must be enforced through suitable boundary conditions of the quantum fields.

Instead of the equilibrium density operator (\ref{densop}), we can use the general 
covariant formula (\ref{covdens2}) as a starting point and look for the 
implications on $\beta, \omega$ and $\xi$ of thermodynamical equilibrium conditions,
i.e. of eqs.~(\ref{divvan}) and (\ref{boundary}); in principle, with this approach,
we could find new forms of equilibrium distributions. For the divergence to vanish,
according to eq.~(\ref{div}), one needs to have:
$$ 
  \partial_\lambda \xi = 0   \qquad \qquad  \partial_\lambda \omega_{\mu \nu} = 0
$$
and, again, the equations (\ref{equilib}). The first of (\ref{equilib})
entails that the inverse temperature four-vector ought to be a Killing vector, a
well known condition for equilibrium \cite{israel2,chrobok}. Together with the second 
of eqs.~(\ref{equilib}), this leads to \cite{degroot}:
\be\label{solution}
  \beta_\mu = b_\mu + \omega_{\mu \nu} x^\nu
\ee
with $b$ constant four-vector. This expression of the inverse temperature four-vector 
comprises all possible forms of relativistic thermodynamical equilibria; the 
rotating case (\ref{betarot}) actually corresponds to $b_\mu = (1/T,{\bf 0})$
and $\omega$ given by eq.~(\ref{omegat}).

By using the eq.~(\ref{covdens2}) one can write down the entropy:
\be 
 S = \log Z + \int_{\Sigma} \di\Sigma_\mu \; \left( T^{\mu \nu} \beta_\nu -
 \frac{1}{2} {\cal S}^{\mu,\lambda \nu} \omega_{\lambda \nu} - \xi \, j^\mu \right) 
\ee
where the symbols without hat denote the mean values of quantum operators (i.e.
$\tr (\wrho \widehat A) = A$) \footnote{It should be pointed out that the mean value 
of operators involving quantum relativistic fields are generally divergent (e.g. 
$T^{00}$ for a free field has an infinite zero point value). To remove the infinities, 
the mean values must be renormalized, what can be simply done for free fields by using
normal ordering in all expressions, including the density operator itself. Henceforth,
it will be understood that all the mean values of operators are the renormalized ones.}. 
For an entropy current to exist in relativistic 
thermodynamics \cite{israel1,israel2} so that the total entropy can be written as 
$S = \int_{\Sigma} \di\Sigma_\mu \; s^\mu$, the logarithm of the partition 
function {\em must} be written as an integral over the same hypersurface $\Sigma$ 
of a vector field $\Phi^\mu$, hereby defined as {\em thermodynamic potential current}:
\be\label{partfun}
 \log Z = \int_{\Sigma} \di\Sigma_\mu \; \Phi^\mu
\ee
\footnote{Note that in this formula $\log Z$ is meant to be the renormalized one, see 
previous footnote} so that the entropy current reads:
\be\label{entr1}
  s^\mu = \Phi^\mu + T^{\mu \nu} \beta_\nu - \frac{1}{2} {\cal S}^{\mu,\lambda \nu} 
   \omega_{\lambda \nu} - j^\mu  \xi 
\ee
In principle, the existence of the thermodynamic potential current could be proved 
working out the trace of $Z \wrho$ in eq.~(\ref{covdens2}), but this requires lenghty
manipulations of the commutators of $\widehat T^{\mu \nu}$ and so we rather {\em assume} 
(\ref{partfun}) invoking the existence of an entropy current. At equilibrium, the 
four-vector field  $\Phi$ must be divergence-free and it must have vanishing outward 
flux through $\partial \Sigma$ for the partition function 
(\ref{partfun}) to be stationary and independent of the space-like hypersurface $\Sigma$.
In view of the (\ref{entr1}), the condition $\partial_\mu \Phi^\mu = 0$ is indeed 
a consequence of the requirement of vanishing entropy production ($\partial_\mu s^\mu 
= 0$) at equilibrium. We point out that the thermodynamic potential current is not 
uniquely defined, as one may add a divergence of an antisymmetric tensor field 
with suitable boundary conditions to obtain the same partition function in 
eq.~(\ref{partfun}). 


The mean value of the stress-energy tensor at equilibrium reads:
\bea
 && T^{\mu \nu}(x) = \tr (\wrho \widehat T^{\mu \nu}) =
  \frac{1}{Z} \tr \left( \widehat T^{\mu \nu}(x) \exp \left[ \int_{\Sigma} 
  \di\Sigma_\mu \; \right. \right. \nonumber \\
 && \left. \left. \left( - \widehat T^{\mu \nu} \beta_\nu + 
  \frac{1}{2} \widehat{\cal S}^{\mu,\lambda \nu} \omega_{\lambda \nu} + 
  \xi \, \widehat j^\mu \right) \right] \right)
\eea
where (\ref{covdens2}) has been used. Let us now fix the space-like hypersurface 
$\Sigma$ and write $\di \Sigma_\mu = \di \Sigma n_\mu$, where $n^\mu$ is its 
normal time-like unit vector. One can obtain a contraction of the mean value of 
the stress-energy tensor with the normal vector $n_\mu$ by taking a functional 
derivative with respect to the four-temperature vector $\beta$ seen as a function 
of $x$, keeping $\omega$ and $\xi$ fixed, with respect to the measure $\di \Sigma$:
\bea\label{fundev1}
 && - n_\mu T^{\mu \nu}(x) = \frac{1}{Z} \frac{\delta}{\delta \beta_\nu(x)}
  \tr \left( \exp \left[ \int_{\Sigma} \di\Sigma \; n_\mu \left( - 
  \widehat T^{\mu \nu} \beta_\nu (x) \right. \right. \right. \nonumber \\
 && \left. \left. \left. + \frac{1}{2} 
    \widehat{\cal S}^{\mu,\lambda \nu} \omega_{\lambda \nu} + \xi 
    \widehat j^\mu \right) \right] \right) \Bigg|_{\omega,\xi} = 
    \frac{\delta \log Z[\beta]}{\delta \beta_\nu(x)}\Bigg|_{\omega,\xi}
\eea
Formally, the above formula can be shown by using the expansion of the exponential
of sum of operators (Zassenhaus formula) and taking advantage of the ciclicity
of the trace. While the left hand side depends on a vector $n_\mu$, which is arbitrary, the 
rightmost hand side is not manifestly dependent on it. In fact, the functional 
derivative of the partition function includes a hidden dependence on the normal 
vector as the functional derivation implies the choice of a measure, hence of a 
hypersurface $\Sigma$ and a corresponding normal vector.

In view of the formula (\ref{partfun}):
\be\label{fundev2}
 \frac{\delta \log Z[\beta]} {\delta \beta_\nu(x)}\Bigg|_{\omega,\xi} 
  = \frac{\delta} {\delta \beta_\nu(x)} \int_{\Sigma} \di\Sigma_\mu \; \Phi^\mu  
  \Bigg|_{\omega,\xi}
\ee
At equilibrium, the thermodynamic potential current depends on the equilibrium values
of $\beta$,$\omega$ and $\xi$. Taking the functional derivative means moving the field
$\beta$ slightly out of equilibrium, i.e. $\beta (x) = \beta_{\rm eq}(x) + \delta 
\beta(x)$, what may introduce dependences of $\Phi^\mu$ on the derivatives of the $\beta$
field. As it is well known from the theory of functional derivation, provided that the 
perturbation $\delta \beta$ is chosen so as to fulfills suitable boundary conditions, 
the (\ref{fundev2}) yields:
\be\label{se0}
 - n_\mu T^{\mu \nu}(x) = n_\mu \left( \frac{\partial \Phi^\mu}{\partial \beta_\nu} -
 \partial_\alpha \frac{\partial \Phi^\mu}{\partial(\partial_\alpha \beta_\nu)}
 + \ldots \right)\Bigg|_{\rm eq}
\ee
where $\ldots$ stands for terms involving higher order derivatives of the $\beta$ field. 
However, since the vector $n_\mu$ is an arbitrary timelike field, the straightforward 
consequence of (\ref{se0}) is:
\be\label{se1}
  T^{\mu \nu} = - \frac{\slashed \partial \Phi^\mu}{\slashed \partial \beta_\nu} 
  \Bigg|_{\rm eq}
\ee
i.e. the stress-energy tensor is minus the Euler-Lagrange derivative, denoted by 
$\slashed \partial$ and defined by eq.~(\ref{se0}), of the thermodynamic potential 
current. 

Similarly, it can be shown that:
\be\label{spin}
  {\cal S}^{\mu,\lambda \nu} = \frac{\slashed \partial \Phi^\mu}{\slashed \partial 
  \omega_{\lambda \nu}} \Bigg|_{\rm eq}   \qquad \qquad 
  j^{\mu} = \frac{\slashed \partial \Phi^\mu}{\slashed \partial \xi} \Bigg|_{\rm eq}
\ee
%


We are now going to work out the above formulae in the simplest instance of thermodynamic 
equilibrium, which is the familiar one with $b_\mu = 1/T_0 u_\mu$, with $u_\mu = const$
normalized four-velocity, $\xi= \mu/T = \mu_0/T_0 = const$, and $\omega=0$, i.e. no rotation. 
Thus $\beta^\mu = b^\mu = 1/T_0 u^\mu$ with $T_0$ the invariant temperature by definition. 
The resulting density operator $\wrho = 1/Z \exp[-\widehat P \cdot \beta + \widehat Q \xi]$ 
is invariant for translations, implying that all mean values of fields (including 
stress-energy tensor) are constant in space-time. The thermodynamic potential current 
$\Phi^\mu$ at equilibrium is a vector function of $\beta$, $\omega$ and $\xi$, but since 
$\beta$ is the only non-vanishing vector field, it can only be of the form:
\be\label{phieq}
  \Phi^\mu = p (\beta^2,\xi) \beta^\mu
\ee
where the physical meaning of the scalar function $p(\beta^2,\xi)$ is to be found.
The above form of the thermodynamic potential current is unambiguous because, due to the 
constancy of the arguments, any additional divergence of an antisymmetric tensor
field vanishes. For this special kind of equilibrium, it was written down first in 
ref.~\cite{israel2}.

In eq.~(\ref{se1}), all terms of the Euler-Lagrange derivative of the $\Phi$ function involving 
derivatives vanish at equilibrium because, e.g.:
\bea
 && \partial_\alpha \frac{\partial \Phi^\mu}{\partial(\partial_\alpha \beta_\nu)}\Bigg|_{\rm eq}
 = \frac{\partial \Phi^\mu}{\partial \beta^\rho \partial(\partial_\alpha \beta_\nu)} 
  \partial_\alpha \beta^\rho \Bigg|_{\rm eq} \nonumber \\
 && + \frac{\partial \Phi^\mu}{\partial \partial_\sigma \beta^\rho \partial(\partial_\alpha 
   \beta_\nu)} \partial_\alpha \partial_\sigma \beta^\rho \Bigg|_{\rm eq} + \ldots
\eea
and since all derivatives of the $\beta$ field vanish at equilibrium, this term 
altogether vanishes. The same applies to possible terms involving derivative of the $\Phi$
function with respect to higher order derivatives of the $\beta$ field. Furthermore,
it is not difficult to realize that the only term of $\partial \Phi^\mu/\partial \beta_\nu$
contributing at equilibrium is the derivative of (\ref{phieq}) itself, that is:
\be\label{se3}
  T^{\mu \nu} = -\frac{\partial \Phi^\mu}{\partial \beta_\nu}\Bigg|_{\rm eq} = 
  - \frac{\partial \Phi^\mu|_{\rm eq}}{\partial \beta_\nu}
\ee
and, similarly:
\be\label{jmu}
  j^\mu = \frac{\partial \Phi^\mu|_{\rm eq}}{\partial \xi} = 
  \frac{\partial p}{\partial \xi}\Bigg|_{\beta^2} \beta^\mu \equiv 
  n u^\mu
\ee
where (\ref{phieq}) has been used. By using (\ref{se3}) and (\ref{phieq}) we get:
\be\label{se4}
  T^{\mu \nu} = - 2 \frac{\partial p}{\partial \beta^2}\Bigg|_\xi \beta^\mu \beta^\nu 
  - p g^{\mu \nu}
\ee
This form would be enough to identify $p$ as the pressure, because if $u = (1,{\bf 0})$ 
then $p$ is the diagonal element of the spacial part of the stress-energy tensor.  
This identification is confirmed by the expression of the proper energy density $\rho$
obtained from (\ref{se4}):
\be\label{gibbs}
   \rho \equiv T^{\mu \nu} u_\mu u_\nu = - 2 \frac{\partial p}{\partial \beta^2}\Bigg|_\xi 
   \beta^2 - p
\ee
and the eq.~(\ref{se4}) turns into the familiar:
\be\label{traditional}
  T^{\mu \nu} = (\rho + p) u^\mu u^\nu - p g^{\mu \nu}
\ee
The equation (\ref{gibbs}), that is $2 \partial p/\partial \beta^2|_\xi = 
- (\rho + p)/\beta^2$ is just an alias of the Gibbs-Duhem relation 
$\partial p/\partial T_0|_{\mu_0} = s$ which can be readily checked taking into 
account that the above derivative is taken by keeping $\xi=\mu_0/T_0$ fixed and 
that $T_0s=\rho + p - \mu_0 n $ (see below). Similarly, the eq.~(\ref{jmu}) is 
an alias of the relation $\partial p/\partial \mu_0|_{T_0} = n$.

Finally, we show that all known thermodynamic relations involving proper entropy 
density are also recovered. From (\ref{entr1}) with $\omega=0$, using (\ref{phieq}) 
we get, by contracting with the four-velocity $u$:
\be\label{entr2}
 s \equiv s^\mu u_\mu = p \sqrt{\beta^2} + \rho \sqrt{\beta^2} - \xi n 
\ee
which, by using $\sqrt{\beta^2} = 1/T_0$ and $\xi = \mu_0/T_0$ reads as the familiar 
relation $T_0 s= \rho + p - \mu_0 n$. We can obtain the differential of entroopy 
current (difference between nearby equilibrium states with $\beta$ and $\xi$ being 
the understood parameters):
\be\label{diff1}
  \di s^\mu = \di \Phi^\mu + T^{\mu \nu} \di \beta_\nu + \beta_\nu \di T^{\mu \nu}
  - \xi \di j^\mu - j^\mu \di \xi
\ee
Since:
\be
  \di \Phi^\mu = \frac{\partial \Phi^\mu}{\partial \beta_\nu} \di \beta^\nu
  + \frac{\partial \Phi^\mu}{\partial \xi} \di \xi
\ee
in view of the (\ref{se3}),(\ref{jmu}), the eq.~(\ref{diff1}) turns into:
\be\label{diff2}
  \di s^\mu = \beta_\nu \di T^{\mu \nu} - \xi \di j^\mu 
\ee
This equation was obtained by Israel and Stewart \cite{israel2} resorting to several
assumptions concerning the familiar form of thermodynamical equilibrium, including the 
expression (\ref{traditional}) itself; in fact, in our method, both (\ref{diff2}) and
(\ref{traditional}) are consequences of the eq.~(\ref{se1}), derived in turn from 
the general form of statistical density operator (\ref{covdens2}), which is more 
economical and transparent. Contracting the above formula with $u_\mu$ we get:
\bea\label{diff3}
 && u_\mu \di s^\mu = \di s - s^\mu \di u_\mu = \di s - (\Phi^\mu + T^{\mu \nu} \beta_\nu 
  - \xi j^\mu) \di u_\mu \nonumber \\
 && =  \sqrt{\beta^2} u_\mu u_\nu \di T^{\mu \nu} - u_\mu \xi \di j^\mu 
\eea
The expression within round brackets is parallel to $u^\mu$ and this makes the whole 
term vanishing as $u^\mu \di u_\mu = (1/2) \di u^2 = 0$. For the same reason, one has
$ u_\mu u_\nu \di T^{\mu \nu} = \di (u_\mu u_\nu \di T^{\mu \nu}) = \di \rho$
and $\xi u_\mu \di j^\mu = \xi \di n$. Therefore, the eq.~(\ref{diff2}) becomes the 
well known:
\be\label{last}
  T_0 \di s = \di \rho - \mu_0 \di n
\ee

We stress that all formulae (\ref{phieq}-\ref{last}) apply to the thermodynamical 
equilibrium without rotation, i.e. with $\omega=0$ in the (\ref{densop}). The
application of (\ref{se1}) to the most general form of equilibrium (\ref{solution}),
and to slightly out-of-equilibrium situations may shed light on a general relation 
between entropy current and the mean values of stress-energy, spin tensors and
the charge current vector. This will be the subject of further studies.



\end{document}